\documentclass[12pt]{article}

\usepackage{mathrsfs}
\usepackage{fullpage}
\usepackage{amsfonts}
\usepackage{graphicx}
\usepackage{mathrsfs}
\usepackage{amsmath}
\usepackage{amssymb}
\usepackage{float}
\usepackage{rotating,color}

\newtheorem{thm}{Theorem}

\def\var{\mathrm {var}}

\def\diag{\mathrm {diag}}
\newcommand{\bm}{\boldsymbol}

\def\tr{\mathrm {tr}}
\def\U{{\bf U}}
\def\A{{\bf A}}

\def\B{{\bf B}}

\def\S{{\bf S}}
\def\R{{\bf R}}
\def\I{{\bf I}}
\def\bmu{{\bm \mu}}
\def\u{{\bf u}}

\def\X{{\bf X}}
\def\x{{\bf x}}

\def\D{{\bf D}}

\def\tr{\mathrm {tr}}

\def\bth{{\bm\theta}}
\def\bms{{\bm\Sigma}}

\def\cp{\mathop{\rightarrow}\limits^{p}}
\def\cd{\mathop{\rightarrow}\limits^{d}}
\def\squarebox#1{\hbox to #1{\hfill\vbox to #1{\vfill}}}
\def\boxit#1{\vbox{\hrule\hbox{\vrule\kern6pt
          \vbox{\kern6pt#1\kern6pt}\kern6pt\vrule}\hrule}}

\begin{document}
\title{\bf Spatial-Sign based High-Dimensional Location Test}
\author{Long Feng and Fasheng Sun \\{\it Northeast Normal University}}
\date{}
\maketitle
\begin{abstract}
In this paper, we consider the problem of testing the mean vector in the high-dimensional settings. We proposed a new robust scalar transform invariant
 test based on spatial sign. The proposed test statistic is asymptotically normal under elliptical distributions. Simulation studies show that our test is very robust and efficient in a wide range of distributions.
\end{abstract}



\section{Introduction}
Assume $\X_1,\cdots,\X_n$ is an independent sample from $p$-variate
distribution $F(\x-\bth)$  located at $p$-variate center $\bth$. We
consider the following one sample testing problem
\begin{align*}
H_0: \bth=\bm 0 ~~~\text{versus}~~~H_1: \bth\not=\bm 0.
\end{align*}
One typical test statistic is Hotelling's $T^2$. However, it can not
be  applied when $p>n-1$ because of the singularity of the sample
covariance matrix. Recently, many efforts have been devoted to solve
the problem, such as Bai and Saranadasa (1996), Srivastava and Du
(2008), Srivastava (2009), Chen and Qin(2010) and Park and Ayyala
(2013). They established the asymptotic normality of their test
statistics under the assumption of diverging factor model (Bai and
Saranadasa 1996). Even this data structure generates a rich
collection of $X$, it is not easily met in practice. Moreover,
multivariate $t$ distribution or mixtures of multivariate normal
distributions does not satisfy the diverging factor model. This
motivates us to construct a robust test procedure.

Multivariate sign or rank is often used to construct robust test
statistics  in the multivariate setting. Especially, multivariate
sign tests enjoy many desirable properties. First, those test
statistics are distribution-free under mild assumptions, or
asymptotically so. Second, they do not require stringent parametric
assumptions, nor any moment conditions. Third, they have high
asymptotic relative efficiency with respect to the classic
Hotelling's $T^2$ test, especially under the heavy-tailed
distributions. However, the classic spatial-sign test also can not
work in the high-dimensional settings because the scatter matrix is
unable to be estimated. Recently, without estimating the scatter
matrix, Wang, Peng and Li (2014) proposed a high-dimensional
nonparametric test based on the direction of $\X_i$, i.e.
$\X_i/||\X_i||$. Even it is workable and robust in high-dimensional
settings, it loses all the information of the scalar of different
variables and then is not scalar-invariant. In practice, different
components may have completely different physical or biological
readings and thus certainly their scales would not be identical.
Srivastava (2009) and Park and Ayyala (2013) proposed two
scalar-invariant tests under different assumption of correlation
matrix. As shown above, they are not robust for the heavy-tailed
distributions. In this paper, we proposed a new robust test based on
spatial sign. We show that it is scalar invariant and asymptotic
normal under some mild conditions. The asymptotic relative
efficiency of our test with respect to Park and Ayyala (2013)'s test
is the same as the classic spatial-sign test with respect to the
Hotelling's $T^2$ test. Simulation comparisons show that our
procedure has good size and power for a wide range of dimensions,
sample sizes and distributions. All the proofs are given in the
appendix.

\section{Robust High-Dimensional Test}
\subsection{The proposed test statistic}
Assume $\{
\X_{1},\ldots,{\X}_{n}\}$ be a independently and
identically distributed (i.i.d.) random samples from $p$-variate
elliptical distribution with density functions $\mbox{det}({\bf
\Sigma})^{-1/2}g(||{\bf
 \Sigma}^{-1/2}({\bf x}-\bm\theta)||)$
where $\bm\theta$'s are the symmetry centers and ${\bf\Sigma}$'s
are the positive definite symmetric $p\times p$ scatter matrices.
The spatial sign function is defined as $U({\bf x})=||{\bf
x}||^{-1}{\bf x}I({\bf x}\neq {\bf 0})$.
In traditional fixed $p$ circumstance, the following so-called
``inner centering and inner standardization" sign-based procedure is
usually used (cf., Section 6 of Oja 2010)
\begin{align}\label{qn2}
Q_n^2=np\bar{\U}^T\bar{\U},
\end{align}
where $\bar{\U}=\frac{1}{n}\sum_{i=1}^n \hat{\U}_i$,
$\hat{\U}_{i}=U(\S^{-1/2}\X_{ij})$, $\S^{-1/2}$ are Tyler's scatter matrix.
$Q_n^2$ is affine-invariant and can be regarded as a nonparametric
counterpart of Hotelling's $T^2$ test statistic by using the spatial-signs instead of the
original observations $\X_{ij}$'s. However, when $p>n$, $Q_n^2$ is
not defined as the matrix $\S^{-1/2}$ is is not available in
high-dimensional settings.

Motivated by Hettmansperger and Randles (2002), we suggest to find a pair of diagonal matrix $\D$ and vector
$\bth$ for each sample that simultaneously satisfy
\begin{align}\label{seq}
\frac{1}{n}\sum_{j=1}^{n} U(\bm\epsilon_{j})=0 ~~\text{and}~~
\frac{p}{n}\diag\left\{\sum_{j=1}^{n}
U(\bm\epsilon_{j})U(\bm\epsilon_{j})^T\right\}=\I_p,
\end{align}
where $\bm\epsilon_{j}=\D^{-1/2}(\x_{j}-\bth)$.
$(\D,\bth)$ can be viewed as a simplified version of Hettmansperger-Randles (HR) estimator without considering the off-diagonal elements of $\S$. We can adapt the recursive algorithm of Hettmansperger and Randles (2002) to solve
(\ref{seq}). That is, repeat the following three steps until
convergence:
\begin{itemize}
\item[(i)] $\bm\epsilon_{j} \leftarrow \D^{-1/2}(\X_{j}-\bth)$,
~~$j=1,\cdots,n_i$;
\item[(ii)] $\bth \leftarrow
\bth+\frac{\D^{1/2}\sum_{j=1}^{n}U(\bm\epsilon_{j})}{\sum_{j=1}^n||\bm\epsilon_{j}||^{-1}}$;
\item[(iii)] $\D \leftarrow p
\D^{1/2}\diag\{n^{-1}\sum_{j=1}^{n}U(\bm\epsilon_{j})U(\bm\epsilon_{j})^T\}\D^{1/2}$.
\end{itemize}
The resulting estimators of location and diagonal matrix are denoted
as $\hat{\bm \theta}$ and $\hat{\D}$. We
may use the sample mean and sample variances as the initial
estimators.

Then, we define the following test statistic
\begin{align*}
R_n=\frac{2}{n(n-1)}\underset{i<j}{\sum\sum} U\left(\hat{\D}_{ij}^{-1/2}\X_i\right)^TU\left(\hat{\D}_{ij}^{-1/2}\X_j\right)
\end{align*}
where $\hat{\D}_{ij}$ are the corresponding diagonal matrix estimator using leave-two-out sample $\{\X_k\}_{k\not=i,j}^n$.

\subsection{Asymptotic results}
We need the following conditions for
asymptotic analysis:
\begin{itemize}
\item[(C1)]
$\tr(\R^4)=o(\tr^2(\R^2))$, where $\R=\D^{-1/2}\bms\D^{-1/2}$;
\item[(C2)] $n^{-2}p^2/\tr(\R^2)=O(1)$ and $\log p=o(n)$;
\item[(C3)] $(\tr(\R^2)-p)=o(n^{-1}p^2)$.
\end{itemize}
Condition (C1) is the same as the condition (4) in Park and Ayyala (2013). To appreciate condition (C2) and (C3), we consider $\tr(\R^2)=O(p)$ (Srivastava and Du 2008; Srivastava 2009). Then, Condition (C2) and (C3) becomes $p=O(n^2)$ and $p/n\to \infty$. Thus, we could allow the dimension being the square of the sample size. To get the consistency of the diagonal matrix, the dimension must diverging faster than the sample sizes.

The following theorem establishes the asymptotic null distribution
of $R_n$.
\begin{thm}
Under Conditions (C1)-(C3) and $H_0$, as $(p,n)\to \infty$,
$R_n/\sigma_n\cd N(0,1)$, where $\sigma_n^2=\frac{2}{n(n-1)p^2}\tr(\R^2)$.
\end{thm}

We propose the following estimators to estimate the trace terms in
$\sigma_n^2$
\begin{align*}
\widehat{\tr(\R^2)}&=\frac{p^2}{n(n-1)}\sum_{i=1}^{n}\sum_{j\not=i}^{n} \left(U(\hat{\D}_{ij}^{-1/2}(\X_{i}-\hat{\bth}_{ij}))^TU(\hat{\D}_{ij}^{-1/2}(\X_j-\hat{\bth}_{ij}))\right)^2
\end{align*}
where $(\hat{\bth}_{ij},\hat{\D}_{ij})$ are the corresponding spatial median and diagonal matrix estimators using leave-two-out sample $\{\X_k\}_{k\not=i,j}^n$.
By Proposition 2 in Feng et al. (2014), $\widehat{\tr(\R^2)}/\tr(\R^2)\to 1$ as $p,n\to \infty$. Consequently, a ratio-consistent estimator of $\sigma_n^2$ under $H_0$ is $\hat{\sigma}_n^2=\frac{2}{n(n-1)p^2}\widehat{\tr(\R^2)}$. And then we reject the null hypothesis with $\alpha$ level of significance if $R_n/\hat{\sigma}_n>z_{\alpha}$, where $z_{\alpha}$ is the upper $\alpha$ quantile of $N(0,1)$.

Next, we consider the asymptotic distribution of $R_n$ under the alternative hypothesis
\begin{itemize}
\item[(C4)] $\bmu^T\D^{-1}\bmu=O(c_0^{-2}\sigma_n)$ where $c_0=E(||\D^{-1/2}(\X_i-\bmu)||)$.
\end{itemize}

\begin{thm}
Under Conditions (C1)-(C4), as $(n,p)\to \infty$,
\begin{align*}
\frac{R_n-c_0^2\bmu^T\D^{-1}\bmu}{\sqrt{\sigma_n^2+\frac{4c_0^2}{np}\bmu^T\D^{-1}\bms\D^{-1}\bmu}} \cd N(0,1)
\end{align*}
\end{thm}
Theorem 1 and 2 allow us to compare the proposed test with some existing work in terms of limiting efficiency. In order to obtain an explicit
expression for comparison use, we assume that $\lambda_{\max}(p^{-1}\R)=o(n^{-1})$ and then $\frac{4c_0^2}{np}\bmu^T\D^{-1}\bms\D^{-1}\bmu=o(\sigma_n^2)$.
Thus, the asymptotic power of our proposed test under the local alternative is
\begin{align*}
\beta_S(\bmu)=\Phi\left(-z_{\alpha}+\frac{c_0^2np\bmu^T\D^{-1}\bmu}{\sqrt{2\tr(\R^2)}}\right).
\end{align*}
In comparison, Park and Ayyala (2013) showed that the asymptotic power of their proposed test (abbreviated as PA hereafter) is
\begin{align*}
\beta_{PA}(\bmu)=\Phi\left(-z_{\alpha}+\frac{n\bmu^T\tilde{\D}^{-1}\bmu}{\sqrt{2\tr(\tilde{\R}^2)}}\right).
\end{align*}
where $\tilde{\D}$ and $\tilde{\R}$ are the variance and correlation matrix of $\X_i$, respectively. Thus, the asymptotic relative efficiency (ARE) of $R_n$ with PA test is
\begin{align*}
{\rm ARE}(R_n, {\rm PA})= \frac{c_0^2p\bmu^T\D^{-1}\bmu}{\bmu^T\tilde{\D}^{-1}\bmu}\sqrt{\frac{\tr(\tilde{\R}^2)}{\tr({\R}^2)}}=&c_0^2E(||\bm \epsilon||^2).
\end{align*}
where the last equality is followed by $\tr(\tilde{\R}^2)=\tr(\R^2)$ and $\tilde{\D}=p^{-1}E(||\bm \epsilon||^2)\D$. Similar to the proof of Theorem 1, under Condition (C3), we can show that $c_0=E(||\bm\epsilon||^{-1})(1+o(1))$. Thus,
\begin{align*}
{\rm ARE}(R_n, {\rm PA})=&E^2(||\bm \epsilon||^{-1})E(||\bm \epsilon||^2)
\end{align*}
If $\X_i$ are generated from multivariate $t$-distribution with $\nu$ degrees of freedom ($\nu>2$),
\begin{align*}
{\rm ARE}(R_n, {\rm PA})=&\frac{2}{\nu-2}\left(\frac{\Gamma((\nu+1)/2)}{\Gamma(\nu/2)}\right)^2.
\end{align*}
Table 1 reports the ARE with different $\nu$. Under the multivariate normal distribution ($\nu=\infty$), our SS test is the same powerful as PA test. However, our SS test is much more powerful than PA test under the heavy-tailed distributions.
 \begin{table}
           \centering
           \caption{ARE($R_n$, PA) with different $\nu$.}
           \vspace{0.1cm}
      \renewcommand{\arraystretch}{0.8}
     \tabcolsep 7pt
         \begin{tabular}{cccccc}\hline \hline
  & $\nu=3$ & $\nu=4$ &$\nu=5$ &$\nu=6$ & $\nu=\infty$ \\
  {\rm ARE} & 2.54 & 1.76 & 1.51 & 1.38 & 1.00  \\ \hline \hline
               \end{tabular}\label{st7}
           \end{table}

In contrast, Wang, Peng and Li (2014) showed that the power of their test (abbreviated as WPL) is
\begin{align*}
\beta_{WPL}(\bmu)=\Phi\left(-z_{\alpha}+\frac{n\bmu^T\A^2\bmu}{\sqrt{2\tr(\B^2)}}\right)
\end{align*}
where $\A=E(||\bm\varepsilon_i||^{-1}(\I_p-U(\bm\varepsilon_i)U(\bm\varepsilon_i)^T))$, $\B=E(U(\bm\varepsilon_i)U(\bm\varepsilon_i)^T)$ and $\bm\varepsilon_i=\X_i-\bmu$. First, if all the diagonal elements of $\bms$ are equal, i.e. $d_i=\delta$, under Condition (C3), we can show that $\A=c_0\I_p(1+o(1))$ and $\tr(\B^2)=p^{-2}\delta^2\tr(\R^2)$. Then,
\begin{align*}
\beta_{WPL}(\bmu)=\Phi\left(-z_{\alpha}+\frac{c_0^2np\bmu^T\D^{-1}\bmu}{\sqrt{2\tr(\R^2)}}\right).
\end{align*}
Thus, our SS test has the same power as WPL test in this case. However, their test is not scalar-invariant. To appreciate the effect of scalar-invariance, we consider the following representative cases. Let $\bms$ be a diagonal matrix. The first half diagonal elements of $\bms$ are all $\tau_1^2$ and the rest diagonal elements are all $\tau_2^2$. The mean only shift on the first half components, i.e. $\mu_i=\zeta, i=1,\cdots,p/2$ and the others are zeros. Thus,
\begin{align*}
\beta_{S}(\bmu)=\Phi\left(-z_{\alpha}+\frac{nE^2(||\bm \epsilon||^{-1})\zeta^2}{2\sqrt{2p}\tau_1^2}\right).
\end{align*}
However, it is difficult to calculate the explicit form of $\beta_{WPL}$ for arbitrary $\tau_1^2, \tau_2^2$. We only consider two special cases. If $\tau_1^2 \gg \tau_2^2$,
\begin{align*}
\beta_{WPL}(\bmu)\approx\Phi\left(-z_{\alpha}+\frac{nE^2(||\bm \epsilon||^{-1})\zeta^2}{2\sqrt{p}\tau_1^2}\right).
\end{align*}
Thus, ARE($R_n$,WPL) has a positive lower bound of $1/\sqrt{2}$. However, if $\tau_2^2\gg \tau_1^2$,
\begin{align*}
\beta_{WPL}(\bmu)\approx \Phi\left(-z_{\alpha}+\frac{nE^2(||\bm \epsilon||^{-1})\zeta^2}{2\sqrt{p}\tau_2^2}\right).
\end{align*}
Then, ARE($R_n$,WPL)=$\tau_2^2/(\sqrt{2}\tau_1^2)$ could be very large. This property shows the necessity of
a test with the scale-invariance property.

\section{Simulation}
Here we report a simulation study designed to evaluate the
performance of the proposed SS test. All the simulation results are
based on 2,500 replications. The number of variety of multivariate
distributions and parameters are too large to allow a comprehensive,
all-encompassing comparison. We choose certain representative
examples for illustration. The following scenarios are firstly
considered.
\begin{itemize}
\item[(I)] Multivariate normal distribution. $\X_i\sim N(\bm\theta,\R)$.
\item[(II)] Multivariate normal distribution with different component variances.
 $\X_i\sim N(\bm\theta,\bms)$, where $\bms=\D^{1/2}\R\D^{1/2}$ and $\D=\diag\{d_{1}^2,\cdots,d_{p}^2\}$, $d_{j}^2=3$, $j\le p/2$ and $d_{j}^2=1$, $j>p/2$.
\item[(III)] Multivariate $t$-distribution $t_{p,4}$.   $\X_{i}$'s are generated from $t_{p,4}$ with $\bms=\R$.
\item[(IV)] Multivariate $t$-distribution with different component variances. $\X_{i}$'s are generated from $t_{p,4}$ with $\bms=\D^{1/2}\R\D^{1/2}$ and $d_{j}^2$'s are generated from $\chi^2_4$.
\item[(V)] Multivariate mixture normal distribution $\mbox{MN}_{p,\gamma,9}$. $\X_{i}$'s are generated from  $\gamma
f_p(\bm\theta,\R)+(1-\gamma)f_p(\bm\theta,9\R)$, denoted by
$\mbox{MN}_{p,\gamma,9}$, where $f_p(\cdot;\cdot)$ is the density
function of $p$-variate multivariate normal distribution. $\gamma$
is chosen to be 0.9.
\end{itemize}
Here we consider the correlation matrix $\R=(0.5^{|i-j|})_{1\le i,j\le p}$. Two sample sizes $n=50,100$ and three dimensions $p=200,400,1000$ are considered. For power comparison, under $H_1$, we consider two patterns of allocation for $\bmu$. One is dense case, i.e. the first $50\%$ components of $\bmu$ are zeros. The other is sparse case, i.e. the first $95\%$ components of $\bmu$ are zeros. To make the power comparable among the configurations of $H_1$, we set $\eta=:||\bmu||^2/\sqrt{\tr^2(\bms)}=0.03$ throughout the simulation. And the nonzeros components of $\bmu$ are all equal. Table 2 reports the empirical sizes and power of SS, PA and WPL tests for multivariate normal (Scenario I and II) and non-normal (Scenario III, IV and V) distributions, respectively.
From Table 2, we observe that our SS test can control the empirical sizes very well in all cases. WPL test can also maintain the significant level very well. However, the empirical sizes of the PA tests is a little larger than the nominal level in many cases, especially for the non-normal distributions.
Under Scenario I and II, PA test has certain advantages over SS as we would expert because the underlying distribution is multivariate normal. However, under the non-normal distributions, our SS test performs significantly better than PA test. It is consistent with the theoretical results in Section 2. When the component variances are same (Scenario I, III and V), the power of our SS test is similar to WPL test. Even we need to estimate the scalar matrix, we do not lose much efficiency in these cases. However, when the component variances are not equal (Scenarios (II) and (IV)), our SS test, even PA test, are much more powerful than WPL test, which further shows that a scalar-invariant test is necessary. All these results show that our SS test is very powerful and robust in a wide range of distributions.

 \begin{table}
                     \centering
                     \caption{Empirical sizes and power ($\%$) comparison at 5\% significance under Scenarios (I)-(V)}
                     \vspace{0.1cm}
                \renewcommand{\arraystretch}{1.0}
               \tabcolsep 7pt{
                   \begin{tabular}{ccccccccccccc}\hline \hline
                   && \multicolumn{3}{c}{Size} && \multicolumn{3}{c}{Dense} &&\multicolumn{3}{c}{Sparse}\\ \cline{3-5} \cline{7-9}\cline{11-13}
 $n$ & $p$ &SS &PA & WPL &&SS &PA & WPL &&SS &PA & WPL\\
 \multicolumn{13}{c}{Scenario I} \\ \hline
   50& 200& 5.4& 6.2& 5.2& &29 &31& 29&&31 &33 &31\\
   50& 400& 6.5& 7.3& 6.6& &29 &33& 30&&31 &34 &31\\
   50&1000& 4.7& 6.9& 5.7& &25 &33& 29&&25 &32 &29\\
  100& 200& 6.2& 6.5& 6.3& &61 &63& 61&&66 &68 &66\\
  100& 400& 5.9& 6.2& 5.1& &63 &64& 63&&67 &68 &68\\
  100&1000& 5.3& 6.2& 5.3& &63 &65& 64&&66 &67 &67\\
   \multicolumn{13}{c}{Scenario II} \\ \hline
   50& 200& 5.6& 6.3& 5.7&& 63& 66& 25&& 70& 72& 29 \\
   50& 400& 6.5& 7.3& 5.4&& 69& 72& 28&& 69& 72& 30 \\
   50&1000& 4.6& 6.9& 6.1&& 67& 74& 28&& 66& 73& 28\\
  100& 200& 5.9& 6.3& 5.0&& 95& 96& 62&& 98& 97& 70\\
  100& 400& 5.9& 6.1& 6.2&& 97& 97& 66&& 98& 98& 71\\
  100&1000& 5.3& 6.2& 6.3&& 98& 98& 67&& 98& 99& 68\\
 \multicolumn{13}{c}{Scenario III} \\ \hline
   50& 200& 5.3& 4.3& 5.2&& 51& 36& 51&& 58& 40& 58\\
   50& 400& 6.4& 6.9& 6.6&& 54& 39& 55&& 57& 41& 58\\
   50&1000& 4.4& 7.1& 5.7&& 51& 37& 56&& 51& 39& 57\\
  100& 200& 6.0& 7.8& 6.3&& 89& 69& 89&& 93& 72& 93\\
  100& 400& 5.9& 6.6& 5.1&& 91& 66& 91&& 93& 71& 93\\
  100&1000& 5.4& 6.1& 5.3&& 93& 69& 93&& 94& 70& 95\\
   \multicolumn{13}{c}{Scenario IV} \\ \hline
   50& 200& 5.3& 4.3& 6.5&& 87 & 67& 50&& 92 & 74& 58\\
   50& 400& 6.4& 6.9& 5.9&& 89 & 74& 56&& 97 & 88& 60\\
   50&1000& 4.4& 7.1& 6.1&& 99 & 88& 57&& 99 & 92& 58\\
  100& 200& 6.0& 7.8& 6.0&& 100& 94& 90&& 100& 93& 94 \\
  100& 400& 5.9& 6.6& 6.2&& 100& 94& 92&& 100& 95& 94\\
  100&1000& 5.4& 6.1& 5.9&& 100& 99& 94&& 100& 97& 96\\
     \multicolumn{13}{c}{Scenario V} \\ \hline
   50& 200& 5.5& 6.0& 5.2&& 44& 34& 44&& 51& 38& 50\\
   50& 400& 6.5& 6.8& 6.6&& 49& 39& 50&& 50& 40& 52\\
   50&1000& 4.6& 7.1& 5.7&& 44& 40& 50&& 45& 39& 50\\
  100& 200& 6.2& 6.6& 6.3&& 84& 67& 84&& 89& 70& 89\\
  100& 400& 5.9& 5.1& 5.1&& 87& 66& 87&& 89& 71& 90\\
  100&1000& 5.3& 5.8& 5.3&& 89& 68& 89&& 90& 71& 91\\ \hline \hline
                    \end{tabular}}\label{st1}
                \end{table}

\section{Appendix}
\subsection{Proof of Theorem 1}
By Tyler's expansion,
\begin{align*}
U&(\hat{\D}_{ij}^{-1/2}\X_i)=U(\D^{-1/2}\X_i+(\hat{\D}_{ij}^{-1/2}-\D^{-1/2})\X_i)\\
&=\U_i-(\I_p-\U_i\U_i^T)(\hat{\D}_{ij}^{-1/2}-\D^{-1/2})\U_i+o_p(n^{-1}).
\end{align*}
Then,
\begin{align*}
&\frac{2}{n(n-1)}\underset{i<j}{\sum\sum}U(\hat{\D}_{ij}^{-1/2}\X_i)^TU(\hat{\D}_{ij}^{-1/2}\X_j)\\
=&\frac{2}{n(n-1)}\underset{i<j}{\sum\sum}\U_i^T\U_j-\frac{4}{n(n-1)}\underset{i<j}{\sum\sum}\U_i^T(\hat{\D}_{ij}^{-1/2}-\D^{-1/2})(\I_p-\U_i\U_i^T)\U_j\\
&+\frac{2}{n(n-1)}\underset{i<j}{\sum\sum}\U_i^T(\hat{\D}_{ij}^{-1/2}-\D^{-1/2})(\I_p-\U_i\U_i^T)(\I_p-\U_j\U_j^T)(\hat{\D}_{ij}^{-1/2}-\D^{-1/2})\U_j\\
&+o_p(n^{-2})\\
\doteq&\frac{2}{n(n-1)}\underset{i<j}{\sum\sum}\U_i^T\U_j+J_{n1}+J_{n2}+o_p(\sigma_n)
\end{align*}
Next, we will show that $J_{n1}=o_p(\sigma_n)$.
\begin{align*}
&\frac{2}{n(n-1)}\underset{i<j}{\sum\sum}\U_{i}^T(\hat{\D}_{ij}^{-1/2}-\D^{1/2})[\I_p-\U_{i}\U_{i}^T]\U_{j}\\
=&\frac{2}{n(n-1)}\underset{i<j}{\sum\sum}\U_{i}^T(\hat{\D}_{ij}^{-1/2}-\D^{1/2})\U_{j}
+\frac{2}{n(n-1)}\underset{i<j}{\sum\sum}\U_{i}^T(\hat{\D}_{ij}^{-1/2}-\D^{1/2})\U_{i}\U_{i}^T\U_{j}\\
\doteq & G_{n1}+G_{n2}.
\end{align*}
Next we will show that $E(G_{n1}^2)=o(\sigma_n^2)$.
\begin{align*}
E\left(G_{n1}^2\right)=&\frac{4}{n^2(n-1)^2}\underset{i<j}{\sum\sum}E\left(\left(\U_{i}^T(\hat{\D}_{ij}^{-1/2}-\D^{1/2})\U_{j}\right)^2\right)\\
=&\frac{4}{n^2(n-1)^2}\underset{i<j}{\sum\sum} E\left(\frac{\left(\u_{i}^T\bms^{1/2}\D^{-1/2}(\hat{\D}_{ij}^{-1/2}-\D^{1/2})\D^{-1/2}
\bms^{1/2}\u_{j}\right)^2}{(1+\u_{i}^T(\R-\I_p)\u_{i})(1+\u_{j}^T(\R-\I_p)\u_{j})}\right)\\
\le & \frac{4}{n^2(n-1)^2}\underset{i<j}{\sum\sum}\bigg\{ E\left(\u_{i}^T\bms^{1/2}\D^{-1/2}(\hat{\D}_{ij}^{-1/2}-\D^{1/2})\D^{-1/2}
\bms^{1/2}\u_{j}\right)^2\\
&+C E\left(\left(\u_{i}^T\bms^{1/2}\D^{-1/2}(\hat{\D}_{ij}^{-1/2}-\D^{1/2})\D^{-1/2}
\bms^{1/2}\u_{j}\right)^2 \u_{i}^T(\R-\I_p)\u_{i} \right)\\
&+C
E\left(\left(\u_{i}^T\bms^{1/2}\D^{-1/2}(\hat{\D}_{ij}^{-1/2}-\D^{1/2})\D^{-1/2}
\bms^{1/2}\u_{j}\right)^2 \u_{j}^T(\R-\I_p)\u_{j}
\right)\bigg\},
\end{align*}
where the last inequality follows by the Taylor expansion. Define
${\bf
H}=\bms^{1/2}\D^{-1/2}(\hat{\D}_{ij}^{-1/2}-\D^{1/2})\D^{-1/2}
\bms^{1/2}$ and then according to Lemma 2 in Feng et al. (2014),  $\tr(E({\bf
H}^2))=o(\tr(\R^2))$ and $\tr(E({\bf
H}^4))=o(\tr(\R^4))=o(\tr^2(\R^2))$ by Condition
(C1). By the Cauchy inequality, we have
\begin{align*}
E\big(\u_{i}^T\bms^{1/2}\D^{-1/2}(\hat{\D}_{ij}^{-1/2}-\D^{1/2})&\D^{-1/2}
\bms^{1/2}\u_{j}\big)^2=p^{-2}\tr({\bf H}^2)=o(p^{-2}\tr(\R^2)),\\
E((\u_{i}^T{\bf H}\u_{j})^2\u_{i}^T(\R-\I_p)\u_{i})&\le (E(\u_{i}^T{\bf H}\u_{j})^4E((\u_{i}^T(\R-\I_p)\u_{i})^2)^{1/2}\\
& \le (p^{-4}\tr(E({\bf H}^4))p^{-2}(\tr(\R-\I_p)^2))^{1/2}\\
&=o(p^{-2}\tr(\R^2)),\\
E((\u_{i}^T{\bf H}\u_{j})^2\u_{j}^T(\R-\I_p)\u_{j})&\le (E(\u_{i}^T{\bf H}\u_{j})^4E((\u_{j}^T(\R-\I_p)\u_{j})^2)^{1/2}\\
& \le (p^{-4}\tr(E({\bf H}^4))p^{-2}(\tr(\R-\I_p)^2))^{1/2}\\
&=o(p^{-2}\tr(\R^2)).
\end{align*}
So we obtain that $G_{n1}=o_p(\sigma_n)$. Similarly, we can show that $G_{n2}=o_p(\sigma_n)$ and then $J_{n1}=o_p(\sigma_n)$. Taking the
same procedure as $J_{n1}$, we can also obtain $J_{n2}=o_p(\sigma_n)$. Moreover, by taking the same procedure to $\u_i^T(\R-\I_p)\u_i$ as $G_{n1}$,
\begin{align*}
\frac{2}{n(n-1)}\underset{i<j}{\sum\sum}\U_i^T\U_j=&\frac{2}{n(n-1)}\underset{i<j}{\sum\sum}
\frac{\u_i^T\bms^{1/2}\D^{-1}\bms^{1/2}\u_j}{\sqrt{1+\u_i^T(\R-\I_p)\u_i}\sqrt{1+\u_j^T(\R-\I_p)\u_j}}\\
=&\frac{2}{n(n-1)}\underset{i<j}{\sum\sum}\u_i^T\bms^{1/2}\D^{-1}\bms^{1/2}\u_j+o_p(\sigma_n).
\end{align*}
Thus,
\begin{align*}
R_n=\frac{2}{n(n-1)}\underset{i<j}{\sum\sum}\u_i^T\bms^{1/2}\D^{-1}\bms^{1/2}\u_j+o_p(\sigma_n).
\end{align*}
Next, we will show that
\begin{align*}
\sqrt{\frac{n(n-1)p^2}{{2\tr(\R^2)}}}\frac{2}{n(n-1)}\underset{i<j}{\sum\sum}\u_i^T\bms^{1/2}\D^{-1}\bms^{1/2}\u_j \cd N(0,1)
\end{align*}
Define $W_{nk}=\sum_{i=2}^kZ_{ni}$ where $Z_{ni}=\sum_{j=1}^{i-1}\frac{2}{n(n-1)}\u_i^T\bms^{1/2}\D^{-1}\bms^{1/2}\u_j$. Let
$\mathcal{F}_{n,i}=\sigma\{\u_1,\cdots,\u_i\}$ be the $\sigma$-field generated by $\{\u_j, j\le i\}$. Obviously, $E(Z_{ni}|\mathcal{F}_{n,i-1})=0$ and it follows that $\{W_{nk}, \mathcal{F}_{n,k}; 2\le k \le n\}$ is a zero mean martingale. The central limit theorem will hold if we can show
\begin{align}\label{clt1}
\frac{\sum_{j=2}^{n}E[Z_{nj}^2|\mathcal{F}_{n,j-1}]}{\sigma_n^2}\cp 1.
\end{align}
and for any $\epsilon>0$,
\begin{align}\label{clt2}
\sigma_n^{-2}\sum_{j=2}^{n}E[Z_{nj}^2I(|Z_{nj}|>
\epsilon\sigma_n|)|\mathcal{F}_{n,j-1}]\cp 0.
\end{align}
It can be shown that
\begin{align*}
\sum_{j=2}^{n}E(Z_{nj}^2|\mathcal{F}_{n,j-1})=&\frac{4}{n^2(n-1)^2}\sum_{j=2}^{n}\sum_{i=1}^{j-1}\u_i^T\bms^{1/2}\D^{-1}\bms\D^{-1}\bms^{1/2}\u_i\\
&+\frac{4}{n^2(n-1)^2}\sum_{j=2}^{n}\underset{i_1<i_2}{\sum^{j-1}\sum^{j-1}}\u_{i_1}^T\bms^{1/2}\D^{-1}\bms\D^{-1}\bms^{1/2}\u_{i_2}\\
\doteq& C_{n1}+C_{n2}
\end{align*}
Simple algebras lead to
\begin{align*}
E(C_{n1})=&\sigma_n^2,\\
\var(C_{n1})=&\frac{16}{n^4(n-1)^4}\sum_{j=1}^{n-1}j^2(E((\u_j^T\bms^{1/2}\D^{-1}\bms\D^{-1}\bms^{1/2}\u_j)^2)-p^{-2}\tr^2(\R^2))
\end{align*}
By Lemma 1 in Feng et al. (2014), $E((\u_j^T\bms^{1/2}\D^{-1}\bms\D^{-1}\bms^{1/2}\u_j)^2)=O(p^{-2}\tr^2(\R^2))$. Thus, $\var(C_{n2})=o(\sigma_n^4)$. Then,
$C_{n1}/\sigma_n^2 \cp 1$. Similarly, $E(C_{n2})=0$ and
\begin{align*}
\frac{\var(C_{n2})}{\sigma_n^2}=\frac{32}{n^4(n-1)^4}\sum_{i=3}^n\frac{i(n-i+1)(i-1)}{2}\frac{\tr(\R^4)}{\tr^2(\R^2)}\cp 0
\end{align*}
implies $C_{n2}=o_p(\sigma_n^2)$. Thus, (\ref{clt1}) holds. It remains to show (\ref{clt2}). Note that
\begin{align*}
\sigma_n^{-2}\sum_{j=2}^{n}E[Z_{nj}^2I(|Z_{nj}|>
\epsilon\sigma_n|)|\mathcal{F}_{n,j-1}]\le
\sigma_n^{-4}\epsilon^{-2}\sum_{j=2}^{n}E[Z_{nj}^4|\mathcal{F}_{n,j-1}].
\end{align*}
Accordingly, the assertion of this lemma is true if we can show
\begin{align*}
E\left\{\sum_{j=2}^{n}E[Z_{nj}^4|\mathcal{F}_{n,j-1}]\right\}=o(\sigma_n^4).
\end{align*}
Note that
\begin{align*}
E\left\{\sum_{j=2}^{n}E[Z_{nj}^4|\mathcal{F}_{n,j-1}]\right\}=\sum_{j=2}^{n}E(Z_{nj}^4)=O(n^{-8})\sum_{j=2}^{n}E\left(\sum_{i=1}^{j-1}\u_j^T\bms^{1/2}\D^{-1}\bms^{1/2}\u_i\right)^4.
\end{align*}
which can be decomposed as $3Q+P$ where
\begin{align*}
Q=&O(n^{-8})\sum_{j=2}^n \underset{s<t}{\sum^{j-1}\sum^{j-1}}E(\u_j^T\bms^{1/2}\D^{-1}\bms^{1/2}\u_s\u_s^T\bms^{1/2}
\D^{-1}\bms^{1/2}\u_j\\
&\times \u_j^T\bms^{1/2}\D^{-1}\bms^{1/2}\u_t\u_t^T\bms^{1/2}\D^{-1}\bms^{1/2}\u_j)\\
P=&O(n^{-8})\sum_{j=2}^n\sum_{i=1}^{j-1}E((\u_j^T\bms^{1/2}\D^{-1}\bms^{1/2}\u_i)^4)
\end{align*}
Obviously, $Q=O(n^{-5}p^{-2}E((\u_j^T\bms^{1/2}\D^{-1}\bms\D^{-1}\bms^{1/2}\u_j)^2))=O(n^{-5}p^{-4}\tr^2(\R^2))=o(\sigma_n^4)$. Define $\bms^{1/2}\D^{-1}\bms^{1/2}=(v_{ij})_{1\le i,j \le p}$.
\begin{align*}
E(\u_{s}^{T}\A_3\u_{t})^4&=E\left(\sum_{i=1}^p\sum_{j=1}^p
v_{ij}u_{si}u_{tj}\right)^4\\
&=\sum_{i_1,\ldots,i_4=1}^p\sum_{j_1,\ldots,j_4=1}^p v_{i_1
j_1}v_{i_2 j_2}v_{i_3 j_3}v_{i_4 j_4}
E(u_{si_1}u_{si_2}u_{si_3}u_{si_4})E(u_{tj_1}u_{tj_2}u_{tj_3}u_{tj_4})\\
&=O(p^{-4})\sum_{i_1,\ldots,i_4=1}^p\sum_{j_1,\ldots,j_4=1}^p v_{i_1
j_1}v_{i_2 j_2}v_{i_3 j_3}v_{i_4 j_4}.
\end{align*}
By the Cauchy inequality, we have
\begin{align*}
\sum_{i_1,i_2,i_3,i_4=1}^p\sum_{j_1,j_2,j_3,j_4=1}^p v_{i_1
j_1}v_{i_2 j_2}v_{i_3 j_3}v_{i_4 j_4} & \leq
\frac{1}{4}\sum_{i_1,i_2,i_3,i_4=1}^p\sum_{j_1,j_2,j_3,j_4=1}^p
(v_{i_1
j_1}^2+v^2_{i_2 j_2})(v^2_{i_3 j_3}+v^2_{i_4 j_4})\\
&=\sum_{i_1,i_2,j_1,j_2=1}^p v_{i_1j_1}^2v^2_{i_2 j_2}
=\left(\sum_{i_1,j_1}v_{i_1j_1}^2\right)^2 =\tr^2(\R^2).
\end{align*}
Thus, $P=O(n^{-6}p^{-4}\tr(\R^2))=o(\sigma_n^4)$.

\subsection{Proof of Theorem 2}
By the Tyler' expansion,
\begin{align*}
U(\hat{\D}_{ij}^{-1/2}\X_i)=&\U_i-(\I_p-\U_i\U_i^T)(\hat{\D}_{ij}^{-1/2}-\D^{-1/2})\U_i\\
&+r_i^{-1}(\I_p-\U_i\U_i^T)\D^{-1/2}\bmu+o_p(n^{-1})
\end{align*}
Then, taking the same procedure as Theorem 1, we have
\begin{align*}
R_n=&\frac{2}{n(n-1)}\underset{i<j}{\sum\sum}\u_i^T\bms^{1/2}\D^{-1}\bms^{1/2}\u_j+\frac{2}{n(n-1)}\underset{i<j}{\sum\sum}r_i^{-1}\U_j^T(\I_p-\U_i\U_i^T)\D^{-1/2}\bmu\\
&+\frac{2}{n(n-1)}\underset{i<j}{\sum\sum}r_j^{-1}\U_i^T(\I_p-\U_j\U_j^T)\D^{-1/2}\bmu\\
&+\frac{2}{n(n-1)}\underset{i<j}{\sum\sum}r_i^{-1}r_j^{-1}\bmu^T\D^{-1/2}(\I_p-\U_i\U_i^T)(\I_p-\U_j\U_j^T)\D^{-1/2}\bmu+o_p(n^{-2}).
\end{align*}
And
\begin{align*}
\frac{2}{n(n-1)}\underset{i<j}{\sum\sum}r_j^{-1}\U_i^T(\I_p-\U_j\U_j^T)\D^{-1/2}\bmu&=\frac{1}{n}\sum_{j=1}^n c_0\u_j^T\bms^{1/2}\D^{-1}\bmu+o_p(\sigma_n),\\
\frac{2}{n(n-1)}\underset{i<j}{\sum\sum}r_i^{-1}r_j^{-1}\bmu^T\D^{-1/2}(\I_p-\U_i\U_i^T)&(\I_p-\U_j\U_j^T)\D^{-1/2}\bmu\\
&=c_0^2\bmu^T\D^{-1}\bmu+o_p(\sigma_n).
\end{align*}
Thus, we can rewrite $R_n$ as follows
\begin{align*}
R_n=&\frac{2}{n(n-1)}\underset{i<j}{\sum\sum}\u_i^T\bms^{1/2}\D^{-1}\bms^{1/2}\u_j+c_0^2\bmu^T\D^{-1}\bmu\\
&+\frac{2}{n}\sum_{j=1}^n c_0\u_j^T\bms^{1/2}\D^{-1}\bmu+o_p(\sigma_n),
\end{align*}
and
\begin{align*}
\var(R_n)=(\sigma_n^2+\frac{4c_0^2}{np}\bmu^T\D^{-1}\bms\D^{-1}\bmu)(1+o(1)).
\end{align*}
Next, taking the same procedure as in the proof of Theorem 1, we can prove the assertion.


\begin{thebibliography}{9}


\bibitem{b}
\textsc{Bai, Z. and Saranadasa, H.} (1996).
Effect of High Dimension: By an Example of a Two Sample Problem.
{\it Statistica Sinica}, {\bf 6},
311--29.

\bibitem{cppw}
\textsc{Chen, L. S., Paul, D., Prentice, R. L. and Wang, P.} (2011),  A Regularized Hotelling's $T^2$ Test for
Pathway Analysis in Proteomic Studies,  {\it Journal of the American Statistical Association}, {\bf 106}, 1345--1360.

\bibitem{cq}
\textsc{Chen, S. X. and Qin, Y-L.} (2010),  A Two-Sample Test for High-Dimensional Data with Applications to
Gene-Set Testing,   {\it  The Annals of Statistics,} {\bf 38}, 808--835.

\bibitem{fzwz} \textsc{Feng, L., Zou, C., Wang, Z. and Zhu, L. X.} (2015),   Two Sample Behrens-Fisher Problem for High-Dimensional Data,  {\it Statistica Sinica}, To appear.

\bibitem{fzw2} \textsc{Feng, L., Zou, C. and Wang, Z.} (2015), Multivariate-sign-based high-dimensional tests for the two-sample location
problem, {\it Journal of the American Statistical Association}, To appear.


\bibitem{gcbl} \textsc{Gregory, K. B., Carroll, R. J., Baladandayuthapani, V. and Lahiri, S. N.} (2014),   A Two-Sample Test For Equality of Means in High
Dimension, , {\it  Journal of the American Statistical Association}, To appear.

\bibitem{gvh} \textsc{Goeman, J., Van De Geer, S. A., and Houwelingen, V.} (2006),  Testing
Against a High-Dimensional Alternative,  {\it Journal of the Royal
Statistical Society, Series B}, {\bf 68}, 477--493.

\bibitem{hh} \textsc{Hall, P. G. and Hyde, C. C.} (1980), {\it Martingale Central Limit
Theory and its Applications}, New York: Academic Press.

\bibitem{hp}  \textsc{Hallin, M. and Paindaveine, D.} (2006),
 Semiparametrically Efficient Rank-based Inference for Shape. I:
Optimal Rank-Based Tests for Sphericity,  \textit{The Annals of Statistics},
{\bf 34}, 2707--2756.

\bibitem{ho} \textsc{Hettmansperger, T. P. and Oja, H.} (1994),  Affine Invariant
Multivariate Multisample Sign Tests,  {\it Journal of the Royal
Statistical Society, Series B,} {\bf 56}, 235--249.

\bibitem{hr} \textsc{Hettmansperger, T. P. and Randles, R. H.} (2002),  A Practical
Affine Equivariant Multivariate Median,  {\it Biometrika}, {\bf 89},
851--860.



\bibitem{mo} \textsc{M$\ddot{\rm o}$tt$\ddot{\rm o}$nen  J. and Oja, H.} (1995),  Multivariate Spatial Sign and
Rank Methods,  {\it Journal of Nonparametric Statistics,} {\bf 5} 201--213.


\bibitem{oj}  \textsc{Oja, H.} (2010), {\it Multivariate Nonparametric Methods with
R}, New York: Springer.

\bibitem{pa} \textsc{Park, J. and Ayyala, D. N.} (2013),  A Test for the Mean Vector in Large Dimension and Small Samples,
{\it Journal of Statistical Planning and Inference}, {\bf 143}, 929-943.



\bibitem{ps} \textsc{Puri, M. L. and Sen, P. K.} (1971), {\it Nonparametric Methods in
Multivariate Analysis}, New York: Wiley.

\bibitem{ra} \textsc{Randles, R. H.} (1992),  A Two Sample Extension of the Multivariate
Interdirection Sign Test,  {\it L1-Statistical Analysis and Related
Methods (ed. Y. Dodge)}, Elsevier, Amsterdam, pp. 295--302.

\bibitem{ra2}  \textsc{Randles, R. H.} (2000),  A Simpler, Affine Invariant, Multivariate,
Distribution-Free Sign Test,  {\it  Journal of the American
Statistical Association}, {\bf 95}, 1263--1268.


\bibitem{sf} \textsc{Serfling, R. J.} (1980), {\it Approximation Theorems of
Mathematical Statistics,} New York: John Wiley \& Sons.


\bibitem{sr1} \textsc{Srivastava, M.S.} (2009),  A Test of the Mean Vector with Fewer Observations than the Dimension under Non-normality,  {\it Journal of Multivariate Analysis}, {\bf 100}, 518--532.

\bibitem{sr2} \textsc{Srivastava, M. S. and Du, M.} (2008),  A Test for the Mean Vector with
Fewer Observations than the Dimension,  {\it Journal of Multivariate Analysis},
{\bf 99}, 386--402.

\bibitem{wpl} \textsc{Wang, L., Peng, B. and Li, R.} (2014).  A High-Dimensional
Nonparametric Multivariate Test for Mean Vector,  Technical report,
the Pennsylvania State University.

\bibitem{zcx} \textsc{Zhong, P., Chen, S. X. and Xu, M.} (2013),  Tests Alternative to
Higher Criticism for High Dimensional Means under Sparsity and
Column-Wise Dependence,  {\it The Annals of Statistics}, {\bf 41},
2703--3110.


\bibitem{zpfw} \textsc{Zou, C., Peng, L, Feng, L. and Wang, Z.}
(2014),  Multivariate-Sign-Based High-Dimensional Tests for
Sphericity,  {\it Biometrika}, {\bf 101}, 229--236.


\end{thebibliography}
\end{document}